\documentstyle[prd,eqsecnum,aps]{revtex}
\def\th{\theta}
\def\a{\alpha}
\def\n{\noindent}
\def\p{\prime}
\def\pp{{\prime\prime}}
\def\t{\times}
\def\b{$^p$}
\def\n{$^d$}

\begin{document}
\draft
\title{Coincidence detection of broadband signals by networks of the
planned interferometric gravitational wave detectors}
\author{ Biplab Bhawal and S.V. Dhurandhar}
\address{Inter University Centre for Astronomy and Astrophysics,
Post Bag 4, Ganeshkhind, Pune-411007, INDIA. \\
E-mail~: biplab@iucaa.ernet.in, sdh@iucaa.ernet.in}
\date{\today}
\maketitle

\begin{abstract}
We describe how the six planned detectors (2 LIGOs, VIRGO, GEO, AIGO,
TAMA) can be used to perform coincidence experiments for the detection
of broadband signals coming from either coalescing compact binaries or
burst sources. We first make  comparisons of the achievable
sensitivities of these detectors under different optical
configurations and find that a meaningful coincidence experiment for
the detection of coalescing binary signals can only be performed by a
network where the LIGOs and VIRGO are operated in power recycling mode
and other medium scale detectors  are operated in dual recycling mode
with a narrower bandwidth. For the model of burst waveform
considered by us (i.e. uniform power spectral density upto 2000Hz), we
find that the relative sensitivity of the power-recycled VIRGO is
quite high as compared to others with their present design parameters
and thus  coincidence experiment  performed by including VIRGO in the
network would not be a meaningful one.
We also calculate the time-delay window sizes by effectively
optimizing the values for different possible networks. The effect of
filtering on the calculation of thresholds has also been discussed. We
then set the thresholds for different detectors and find out the
volume of sky that can be covered by different possible networks and
the corresponding rate of detection of coalescing binaries in the
beginning of the next century. We note that a coincidence experiment
of power-recycled LIGO detectors and VIRGO and dual-recycled GEO and
AIGO can increase the volume of the sky covered by 3.2 times as
compared with only the power-recycled LIGO detectors - (thus
proportionately increasing the event rate) - and by 1.7 times the sky
covered by the power-recycled LIGO-VIRGO network. For the detection of
burst sources of a given energy, we find that a network of
power-recycled LIGOs with dual-recycled GEO and AIGO can increase the
volume of the sky covered by 4.9 times as compared with only the
power-recycled LIGOs.
These values for both the cases are of course far less than the range
that can be covered by only the LIGO-VIRGO network with dual recycling
operation at a later stage, but the accuracy in the determination of
direction, distance and other source parameters will be much better in
a coincidence experiment in which other detectors and especially AIGO
take part.
\end{abstract}


\section{ Introduction}
\label{one}

The direct detection of gravitational waves is perhaps the most
challenging problem in experimental physics today. Efforts initiated
by Weber \cite{Web}  with bar detectors and followed up by several
individuals and research groups in various directions\cite{Kip,Kip1}
over the last three decades towards achieving this goal are now
culminating into the planning and setting up of long baseline laser
interferometric gravitational wave detectors in several countries
around the globe. The prototypes of these already exist in Germany,
Great Britain, Japan and USA.

We can expect that at the turn of the century we shall have a network
of three long baseline laser interferometric gravitational wave
detectors as a result of the successful accomplishment of the American
LIGO (4 Km length)\cite{Ab} and French-Italian VIRGO(3 Km)
\cite{Virgo} projects. Two more detectors, GEO (600m) \cite{Geo} in
Germany
and TAMA (300m) in Japan are under construction. Also, a 500m
detector, AIGO is in the planning stage in Australia.

The most promising sources to be detected by the planned long baseline
interferometers are the compact coalescing binary systems which emit a
typical {\it chirp} like waveform. Because of their inherent broadband
nature, interferometric detectors can be used to follow the changing
frequency of this wave emitted during the inspiral of compact binary
systems during the final stages of their evolution. The rate of such
coalescences is estimated to be about three per year out to a distance
of 100-200 Mpc \cite{Phin}. The problem of detection of the
gravitational wave signal from a coalescing compact binary and
estimation of its parameters for a single detector has been studied by
a number of authors
\cite{Ccb,BSS,SVD,FCher,KLM,K}

For confirming the detection of gravitational waves it is essential to
conduct coincidence observation by checking the consistency of the
responses of two or more detectors at different sites to a
gravitational wave event to eliminate rare and unmodelled sources of
noise in single detectors.
Also, once the instruments achieve their first objective of detecting
the gravitational waves, then the whole network can serve as a
powerful astronomical observatory providing information complementary
to that obtained by means of electromagnetic wave observations
\cite{Schu2,Schu3,Schu4,KS}.
An important problem called the inverse problem  is to determine from
the  response function obtained by means of linear filtering of the
data in each detector the astrophysically important parameters, i.g.,
the distance to the binary, its orientation and position in the sky,
masses and spins. Solutions to this problem have been arrived at for
detector networks,  both in absence of noise \cite{DhTin} and in
presence of noise \cite{Gur,Jar}.

Since a network of three detectors provides three amplitudes and two
independent time delays, it is the minimum configuration needed to
determine the five parameters of the gravitational waves, namely, two
angles representing direction, two independent polarizations, $h_+$
and $h_\times$ in the geocentric coordinate system and the orientation
of polarization angles on the sky, $\psi$. Two independent relative
time delays, in turn, however, yield two source directions. So, a
complete solution for the gravitational wave requires observation by
four laser interferometric broadband detectors.  This provides another
advantage over three detector network: in coincidence experiments with
four detectors, the solution for a transversely polarized quadrupolar
wave will be overdetermined
and thus any inconsistency  among the data would be evidence for other
polarizartion states\cite{Gur}. One can, therefore, test Einstein's
predictions regarding gravitational wave polarization using such a
network.

The responses, $h_+(t)$ and $h_\times (t)$ given by the
interferometers to a source can be calculated as a function of the
absolute distance and other parameters of the binary system. So, by
measuring these responses it is possible to calculate  the absolute
distance of the source
\cite{Schu2,Schu3,Schu4,Gur,Jar,Vschu,CutF}, the orientation of the
orbital plane (but only with an order four symmetry)\cite{Vschu} and
the masses and spins of two bodies\cite{CutF}. As pointed out by
Schutz\cite{Schu3,Schu4,KS} and later studied in detail by
Markovi\'c\cite{Mark}, the determination of the distance of the binary
will also lead to an accurate measurement of the Hubble's constant and
other cosmological parameters of our universe.

In this paper, however, we re-address the problem of the detection of
broadband gravitational waves from the coincidence observation of the
possible networks of the planned interferometric detectors and set the
thresholds for the planned detectors for the observation of both burst
and {\it chirp} waveform from compact coalescing binaries. So, we
consider simple threshold crossing experiments, in which coincidences
are sought between times when the detector outputs cross preset
thresholds.  This problem was first discussed by Schutz\cite{Shu}, but
here we investigate how exactly one should take into account different
sensitivities of the planned detectors and time delays between them to
formulate an optimal way to set the thresholds.

The study of the coincidence experiment among the planned detectors is
very important also due to the following reason: if the source lies
anywhere near the plane formed by the LIGO/VIRGO network, the
accuracies of their direction measurements and the ability to monitor
more than one waveform will suffer a lot\cite{Kip1}. The coincidence
experiment with a fourth detector at a site far out of this plane is
very important to get a good sky coverage
and for the accurate determination of source parameters. The sites
chosen for the Australian detector, AIGO and the Japanese detector,
TAMA fit nicely into these criteria. We now investigate here, by
taking other related factors  into consideration, like sensitivities,
how this network of the planned detectors would turn out to be useful.

We note here that the thresholds for the detectors in a particular
network are to be set at a level which will guarantee with some level
of confidence that any collection of events above the thresholds of
all the detectors will be free from contamination of `false
alarms'. The calculation to determine threshold levels for any
coincident experiment should take into account the following three
considerations \cite{Shu}:
\begin{description}
\item{\rm (a)} {\it Difference in sensitivity of the detectors}: If
${\cal T}_1$ is the threshold of detector 1 which we consider to be
the reference one, then, in order to detect the same gravitational
wave, one should set the threshold of detector $\j$ at $\zeta^\j {\cal
T}_1$, where $\zeta^\j$ is the ratio of the signal to noise ratios of
detectors $\j$ and 1 for gravitational waves from a particular kind of
source. So, the false alarm rate for the coincidence experiment is
kept down mainly by lowering the rate of the noise-generated
accidental events in the more sensititive detector.
\item{\rm (b)} {\it Time-delay window size}: If two detectors are
separated by some distance, then events generated by noise within a
certain time-delay window between detectors will contribute to the
false alarm rate. If ${\cal F}$ is the chosen value of the joint rate
of false alarm to which we want to set our detectors and if the
maximum time delay between the detectors is $W/2$ measurement
intervals, then, for calculation of the threshold, the appropriate
probability to use is ${\cal F}/W$, since each possible event has to
be compared with W possible coincident ones in the other.

\item{\rm (c)} {\it The effect of filtering the data on the setting of
thresholds}: Filtering is an important aspect for the detection  of
signals from coalescing binaries, whereas in case of short duration
bursts the detection process has to depend on the time-series
observation.
\end{description}

In section \ref{second}, we give a short description of the planned
laser interferometric detectors and present the values of various
important parameters used in the design of these interferometers and
make comparisons of their sensitivities for coalescing compact
binaries and millisecond burst sources.
Section \ref{third} describes our calculation of the optimum window
size for networks with detectors at three or more different sites.
In section \ref{4th} we discuss the effect of filtering on setting
thresholds for the observation of coalescing binaries and make a
conservative estimate of the number of filters to be chosen for our
calculation. Finally, we set the thresholds for detectors in different
networks for the observation of two kinds of sources mentioned above
and also present values of range as well as the source rate that can
be achieved by these networks. In section \ref{6th} we summarize our
results and make some concluding remarks.

Unless stated otherwise, we work with units in which $G=c=1$, so that
all quantities are expressed in units of seconds. Sometimes, for
convenience, we express masses in unit of solar mass, the conversion
factor being $1M_\odot=4.926\times 10^{-6}$ sec.

\section{ Description of Detectors and their relative sensitivities}
\label{second}

Table \ref{t0} provides a short description of the site and
configuration of the planned detectors. The letter written in the
second column represents the corresponding detector throughout our
analysis. We, however, use the letter H to denote the two LIGO
detectors at Hanford together. It should also be noted that the 300m
TAMA (Tokyo Advanced Medium-scale Antenna) has been designed to work
as a prototype in its initial stage. The minimum length of the
proposed AIGO detector has been set to be 500m, however, this may
change, but not much, depending on the budget available \cite{San}. It
may be noted that AIGO is the only detector to be built up in the
southern hemisphere.

In the analysis that follows we {\it only} consider the initial LIGO
detectors and {\it not} the advanced ones. One of our main objectives
here is to analyse how well the network of the six detectors can be
used for the detection of gravitational waves in near future. The
global network may completely change by the time the advanced
detectors come up and then a similar analysis can again be repeated
taking into account higher sensitivities and perhaps changed
time-delay windows with newer detectors.

 The interferometer, in essence, is a rather elaborate transducer from
optical path difference to output power. So, by monitoring the change
in output power, it would be possible to detect the changing curvature
of spacetime induced by the passage of a gravitational wave. To
increase the storage time of laser light in arms, either delay lines
(DL) \cite{Wink} or the Fabry Perot (FP) \cite{Dre} cavity system has
to be adopted, both of which have certain advantages and disadvantages
associated with them. However, the advantages of having a smaller
vacuum system and the possibility of accomodating more than one
detector in the same enclosure have led to the choice of the FP system
in the long baseline interferometers.

 In order to reduce the shot noise level, an integral feature of these
detectors will be the use of a high power laser, in conjunction with
variants of optical technique known as light recycling. The first one
of these techniques is called {\it power recycling} \cite{Rec,FPDL},
in which, at a dark fringe operation of the interferometer, the
outgoing laser light is recycled back into the interferometer by
putting a mirror in front of the source, thus enhancing the laser
power. When a gravitational wave passes through an interferometer, it
modulates the phase of the laser light, thus producing sidebands which
travel towards the photodetector\cite{Vin}. So, another variant of
optical technique called {\it dual recycling} incorporates a second
mirror called the {\it signal recycling mirror} placed in front of the
photodetector. This arrangement can store signal sidebands for
sufficiently long times to allow optimum photon noise sensitivity
within a restricted bandwidth \cite{Dual}.
If one makes one of the sidebands resonant with the three mirror
signal recycling cavity (two end mirrors and the signal recycling
mirror), the arrangement is called {\it the narrowband mode of
operation}. If the same recycling cavity is made to be resonant with
the laser light, that is called the {\it broadband mode of
operation}. Wider the band, lesser the sensitivity\cite{KLM,Dual}. One
may adjust this by choosing suitable values of the reflectivity of the
signal recycling mirror.

In our analysis we consider two kinds of sources: millisecond bursts
and the {\it chirp} waveform from coalescing compact binaries. The
Fourier transform of the Newtonian {\it chirp} waveform, under the
stationary phase approximation, is given as
\begin{equation}
{\tilde h}(f)\simeq \cases {
Q{\cal D}^{-1} {\cal M}^{5/6}f^{-7/6}\exp[i\Psi(f)],& for $f_u\ge f> 0$,\cr
{\tilde h}^*(-f)& for $-f_u\le f< 0$, \cr
 0 & for $\vert f\vert>f_u$,\cr}
\label{stph}
\end{equation}
where the phase
\begin{equation}
\Psi(f) = 2\pi f t_c-\phi_c-{\pi\over 4}+{3\over 4}(8\pi {\cal M} f)^{-5/3}.
\label{ph}
\end{equation}
and $f_u$ represents the approximate value of the frequency at which
the inspiral waveform will {\it shut off}. This happens at an orbital
radius of
approximately $6.0M$ where $M$ is the total mass of the binary. So,
$f_u$ is given by
\begin{equation}
f_u = (6^{3/2}\pi M)^{-1}.
\label{fu}
\end{equation}

The function $Q(\th ,\phi ,\psi ,\i )$ is dependent on the direction
of the source, $(\th, \phi)$, the inclination angle of binary, $\i$
and the polarization angle, $\psi$. The {chirp mass} of the binary
($=\mu^{3/5}M^{2/5}$) is denoted by ${\cal M}$ and $\mu$ is the
reduced mass. The quantities, ${\cal D}$ and $t_c$ represent the
distance of the binary and the {\it collision time} of the system (at
which the orbital radius goes to zero) respectively and $\phi_c$ is
the phase of the waveform as it approaches the collision time, $t_c$.

The signal to noise ratio(SNR) squared can be calculated to be\cite{FCher,CutF}
\begin{equation}
(S/N)^2(f) = 4{Q^2 Q_\j^2\over D^2}{\cal M}^{5/3}\int_{f_l}^{f_u}\:
\bigg[f^{7/3} S_h(f)\bigg]^{-1}\, df,
\label{snr}
\end{equation}
where the additional factor, $Q_\j$ is inserted to take into account
the dependence of the detected waveform on the orientation of the
detector, $\j$,
since we will be considering a network of detectors which, in general,
would have different orientations.
We will discuss about our choice of the lower and upper frequency
cut-offs, $f_l$ and $f_u$ after the Eq.(\ref{QU}) below. We also make
an important assumption here that the noise of the detector is
described by a Gaussian density distribution. The noise power spectral
density, $S_h(f)$ is given by
\begin{equation}
S_h(f)=S_h^{\rm seis}(f)+S_h^{\rm susp}(f)+S_h^{\rm int}(f)+
S_h^{\rm shot}(f)+S_h^{\rm quan}(f).
\end{equation}
All the terms of the above expression are discussed below.

The thermal noise has contribution mainly from two sources, the
pendulum suspension and internal vibration of the test masses. These
are given by Eqs.(4.3) and (4.4) respectively in Finn and
Chernoff\cite{FCher}
(for detailed discussion, see Saulson\cite{Saul90} and the analysis by
VIRGO project\cite{Virgo}) ignoring both torsional and violin modes
which contribute to the interferometer signal directly only if the
resonant optical mode is misaligned from the central axis of the
mirror (see Gillespie and Raab\cite{GR}) :
\begin{equation}
S_h^{\rm susp}(f)={2k_B Tf_0\over \pi^3 m Q_0L^2\bigg[(f^2-f_0^2)^2 +
(ff_0/Q_0)^2\bigg]},
\end{equation}
\begin{equation}
S_h^{\rm int}(f)={2k_B Tf_{\rm int}\over \pi^3 m Q_{\rm
int}L^2\bigg[(f^2-f_{\rm int}^2)^2 + (ff_{\rm int}/Q_{\rm
int})^2\bigg]},
\label{intv}
\end{equation}
 where $T$, $m$, $Q_0$ and $f_0$ are the temperature, mass, suspension
quality factor and resonant frequency of the test mass. The values of
$T$ and $f_0$ are taken to be same for all the detectors, 300$^0$K and
1 Hz respectively. The test mass fundamental mode and quality factor
are represented by $f_{\rm int}$ and $Q_{\rm int}$ respectively.

It may be noted that only the fundamental vibrational mode has been
considered in Eq.(\ref{intv}) to evaluate the thermal noise due to the
internal vibration. A more accurate estimation should include more
number of modes to predict the Brownian motion of the mirror surface
(see the detailed analysis by Gillespie and Raab\cite{GR95}). However,
the internal thermal noise
will not be a serious contributing factor to the overall noise curve
of the first LIGO interferometers; but it is likely to be a serious
contributor in advanced interferometers. Typically, more modes can
contribute to thermal noise for larger mirrors and smaller laser beam
spot sizes\cite{GR95}.

Both the above expressions have been obtained by assuming that the
noise would be due to viscous damping rather than structural
damping. Works by various groups\cite{Virgo,Saul90,GR,GR95} suggested,
but did not prove that the primary dissipative force may be due to a
phase lag between the stress and strain in both the pendulum
suspension and the internal vibrations of the test masses.
But this will not be an important issue when applied to the initial
interferometers and thus such an assumption will not affect our
results in this paper.

The idealized expressions for the photon shot noise for FP and DL type
interferometers using power recycling are given by Krolak, Lobo and
Meers\cite{KLM,K}, Eqs.(3.5) and (3.6) :
\begin{equation}
S_h^{\rm shot}(f)\vert_{\rm FP}={\hbar\lambda\over\eta I_0}{A^2\over
\ell}f_c\Bigg[1+
\bigg({f\over f_c}\bigg)^2\Bigg],
\end{equation}
\begin{equation}
S_h^{\rm shot}(f)\vert_{\rm DL}={\hbar\lambda\over\eta I_0}{A^2\over
\ell}f_d{(f/f_d)^2\over\sin^2(f/f_d)},
\end{equation}
where $I_0$, $\lambda$ are laser power and wavelength, $\eta$ is the
quantum efficiency of photodiode, $A^2$ represents mirror losses,
$\ell$ is the detector arm length. The quantity $f_c$ represents the
frequency at which the detector sensitivity has its optimum shot-noise
limited values and is called the {\it recycling knee frequency}. It
can be adjusted by choosing appropriate value for the input mirror
reflectivity of the cavity:
\begin{equation}
f_c={(1-\sqrt{R_1R_2})c\over 4\pi \ell},
\end{equation}
$R_1$ and $R_2$ being the intensity reflectivities of the input mirror
and the end mirror respectively. Similarly, in DL antennas, the
minimum noise spectral density corresponds to a frequency $\simeq 1.2
f_d$ and $f_d$ can be adjusted by choosing the number of bounces, $N$
for a DL system:
\begin{equation}
f_d = {c\over \pi N \ell}
\end{equation}
For the dual recycling mode of operation, where one can work with a
narrower bandwidth, $\Delta f$ around some properly selected
frequency, $f_n$ by adjusting the reflectivity of the signal recycling
mirror, the photon counting noise has the following expression for
both FP and DL system:
\begin{equation}
S_h^{\rm shot}(f)\vert_{\rm dual}={4\hbar\lambda c\over\pi\eta
I_0}\bigg({A^2\over \ell}\bigg)^2\Bigg[1+4\Bigg({f-f_n\over\Delta
f}\Bigg)^2\Bigg].
\end{equation}
It should be noted at this point that the expressions presented above
provide only some idealized numbers for the values of the shot
noise. In reality the situation would be somewhat worse due to the
phase modulation technique\cite{mod} to be incorporated in these
interferometers. The phase modulation technique is necessary in
practice to overcome the effects of imperfect fringe visibility and to
move the signal away from low frequencies where the laser light tends
to have excess noise.

Also, the contribution of the quantum uncertainty in the determination
of the periodic motion of the end masses is given by (see Eq.(121),
Ref.\cite{Kip})
\begin{equation}
S_h^{\rm quan}(f)={8\hbar\over 4\pi^2f^2\ell^2m}.
\label{QU}
\end{equation}

So far we have not considered the seismic noise.  The amplitude of the
seismic noise depends on local characteristics of the seismic
displacement noise at the site as well as the nature of the seismic
isolation circuit and, obviously, will be different for different
detectors (see Saulson\cite{Saul84} and Eq.(4.6) of
Ref.\cite{FCher}). The seismic noise spectrum falls very steeply
($\sim f^{-24}$) and is like a `wall' near a certain frequency which
we call $f_l$. So, practically for $f>f_l$, the seismic noise
contribution is negligibly small as compared to the thermal noise and
the shot noise, while for $f<f_l$, $S_h^{\rm quan}(f)\to\infty$. We
uniformly choose $f_l$ to be 70 Hz for all the detectors for the
purpose of data analysis, however, in reality, this may vary. If the
seismic noise characteristics for all the detectors get properly
determined, then, of course, one can incorporate this noise in the
expression for $S_h(f)$ and may choose a much lower value for $f_l$ to
calculate the relative SNRs to be defined in Eq.\ref{chi} below.

We checked numerically that, if the seismic noise is considered to be
negligible above the chosen value of $f_l$, then according to the
calculation done by us, variation in the choice of $f_l$ within the
range (10Hz - 100Hz) does not change the value of the relative SNR
much (as defined in the following Eqs.\ref{chi} and \ref{chibur}) for
all the detectors in their initial stage.
 We also choose $f_u=2000$ Hz in our calculation of relative SNRs
which correspond to a total mass,$M=2.2 M_\odot$ of the binary
system.

We may note that although we have neglected the seismic noise above a
certain frequency, $f_l$, we have not neglected the quantum noise,
$S_h^{\rm quan}(f)$
which is also quite small there. The reason is that above frequencies
much greater than the pendulum frequency, $f_0$ (=1Hz), the seismic
noise falls as steeply as $f^{-24}$, whereas the quantum noise falls
as $f^{-2}$. So, the latter may become important in a narrower-band
dual recycling operation with a low value of $f_n$ and may act as the
fundamental limit to the sensitivity of the narrow band operation. For
a power recycling operation, of course, both may be neglected above
$f_l$.

Now, referring to the expression for SNR given in Eq.\ref{snr}, we can
define the relative signal to noise ratio of any detector $\j$
($\equiv$ L, H4, H2, V, G, T or A) for a particular binary system by
normalising its own value with that of any of the LIGO full-length
detectors:
\begin{equation}
\zeta^\j_{\rm cb}=\Bigg[ {F_{\rm cb}^{\j}\over F_{\rm cb}^L}\Bigg]^{1/2}
\label{chi}
\end{equation}
where the subscript `cb' represents coalescing binary and
\begin{equation}
F_{\rm cb}^{\j}=\int_{f_l}^{f_u}\: \bigg[f^{7/3} S_h^\j (f)\bigg]^{-1}\, df
\end{equation}
and $S_h^\j (f)$ represents the noise power spectral density of the
detector $\j$. In this first analysis we have ignored the orientation
factors of the detectors, that is in effect we have taken $Q_\j$ to be
same for all the detectors. Statistically, an averaging effect would
be present if we assume isotropic distribution of sources.

To arrive at some close-to-real values of the relative SNRs for the
burst sources, in the absence of our knowledge of the waveform, we
assume the power of the burst to be distributed uniformly
in a frequency range between 70 Hz and 2000 Hz; The level of this
power spectrum is different for different sources. Then, for any
particular burst source, the relative SNRs as normalized to that of a
full-length LIGO can be given as
\begin{equation}
\zeta^\j_{\rm bur}=\bigg[ {F^{\j}_{\rm bur}\over F_{\rm bur}^L}\bigg]^{1/2},
\label{chibur}
\end{equation}
where
\begin{equation}
F_{\rm bur}^{\j}=\int_{f_l}^{f_u}\: \big[ S_h^\j (f)\big]^{-1}\, df
\end{equation}

In Table \ref{t10} we present values of the different parameters we
used in our numerical calculation to arrive at the values of the
relative SNRs, $\zeta^\j_{\rm cb}$ and $\zeta^\j_{\rm bur}$. In most
of the cases, especially for VIRGO\cite{VIN}, GEO\cite{Win},
AIGO\cite{San} and TAMA\cite{FU}, these have been obtained by personal
communication with people involved in designing the planned
detectors. The values for LIGO have been taken from
Ref.\cite{Ab}. Some of these are quoted as {\it conservative
estimates}, while others are quoted to be {\it hopefully achievable
ones}. A few of the values are assumed by us. The value of $R_1$ for
H2 has been chosen such that its cavity finesse becomes approximately
twice that of H4\cite{dhs}. This leads to the same value of recycling
knee frequency for both the detectors.

Table \ref{t20} presents the values of $\zeta^\j_{\rm cb}$ and
$\zeta^\j_{\rm bur}$
obtained by us for both power and dual recycling mode of operation.
{\it We should note that all values, including those obtained for the
dual recycling operation are normalised with respect to the SNR of
power recycled full-length LIGO for the same kind of source}. The dual
recycling parameters are, in all cases, chosen to be those for a
bandwidth of 300Hz around 250Hz as will be used in GEO\cite{Win}. It
should be noted that for different detectors, maximum gain may be
obtained for different values for bandwidth and central
frequency\cite{bw}. Although the dual recycling values for LIGO, VIRGO
and TAMA have been calculated, they do not have any plan for
implementing this in their initial interferometers.

The values of the SNR of VIRGO for the burst waveform in
Table\ref{t20} may seem to be too high as compared to those for other
detectors. As we guess, the reason for such a high value may be the
choice of a high value for the recycling knee frequency, $f_c$, ($\sim
620$ Hz) as compared to that of LIGO ($\sim$ 92 Hz), which leads to a
lower value for VIRGO of the noise at high frequencies (where it is
shot-noise-limited). Since the power of the burst waveform is assumed
by us to be uniformly distributed in a frequency range 70-2000 Hz, the
SNR as calculated for VIRGO turns out to be more.

We have also checked that, as mentioned above, if the seismic noise is
considered to be negligible, the value of $F_{\rm cb}^\j$ (SNR is
proportional to the square root of this) does not change much as one
goes to lower values of $f_l$. For full-length LIGO with power
recycling, this value
increases by only 3.6\% as the value of $f_l$ is reduced from 70Hz to
10Hz and
reduces by 6\% as $f_l$ is increased from 70Hz to 100Hz (the value of
$F_{\rm cb}^L$ for $f_l=70$Hz is $5.01\times 10^{42}$). The variations
in the relative SNRs of other detectors are even less. The reason for
this is the high value of thermal noise at the low frequency range for
all the detectors at their initial stage, which prevents the SNR to
grow much. However, the seismic noise and the choice of $f_l$ will be
very important once the thermal noise is reduced to a sufficient
extent.

\section{ Determination of the window size}
\label{third}

In this analysis, we confine our attention typically to bursts of
duration 1 msec. Then we must sample the noise in the output
effectively 1000 times per second and should set the thresholds of the
detectors such that the {\it false alarm rate}, ${\cal F}$ for the
coincident observation does not go beyond {\it 1 measurement per
year}.

We differ from Ref.\cite{Shu}, while calculating the time delay window size
for three or more detectors. If $W/2$ is the maximum of the values of
time delay between the detectors, then following that analysis, one
can overestimate the window area as the area of the square, $W^2$. We
optimize this case and obtain a much reduced window area by
calculating the possible range of the window length between 3rd and
1st detector by fixing each time the window length between 2nd and
1st. The procedure we follow is explained in detail below. The method
of optimization followed by us gives its best results for networks of
four or more number of detectors.

Since in this analysis we are mainly interested in the problem of
detection by coincidence experiments and not in the more involved
problem of determination of the exact direction and other parameters,
we choose to work with a coordinate system which is fixed with respect
to a particular detector network. The detectors have different
orientations on the surface of the earth and also the waves will
arrive from random directions with random polarization angles.  The
antenna pattern and the coincidence probability for a detector network
averaged over all possible thresholds was computed by Tinto\cite{Tin}
using Monte Carlo methods and integrating over all the random
parameters. In this paper, however, we are mainly concerned about
setting thresholds for the detection of individual events and,
therefore, to avoid complications, we completely ignore the fact that
different detectors would respond differently to a given gravitational
wave.

\subsection{window size for three detectors}
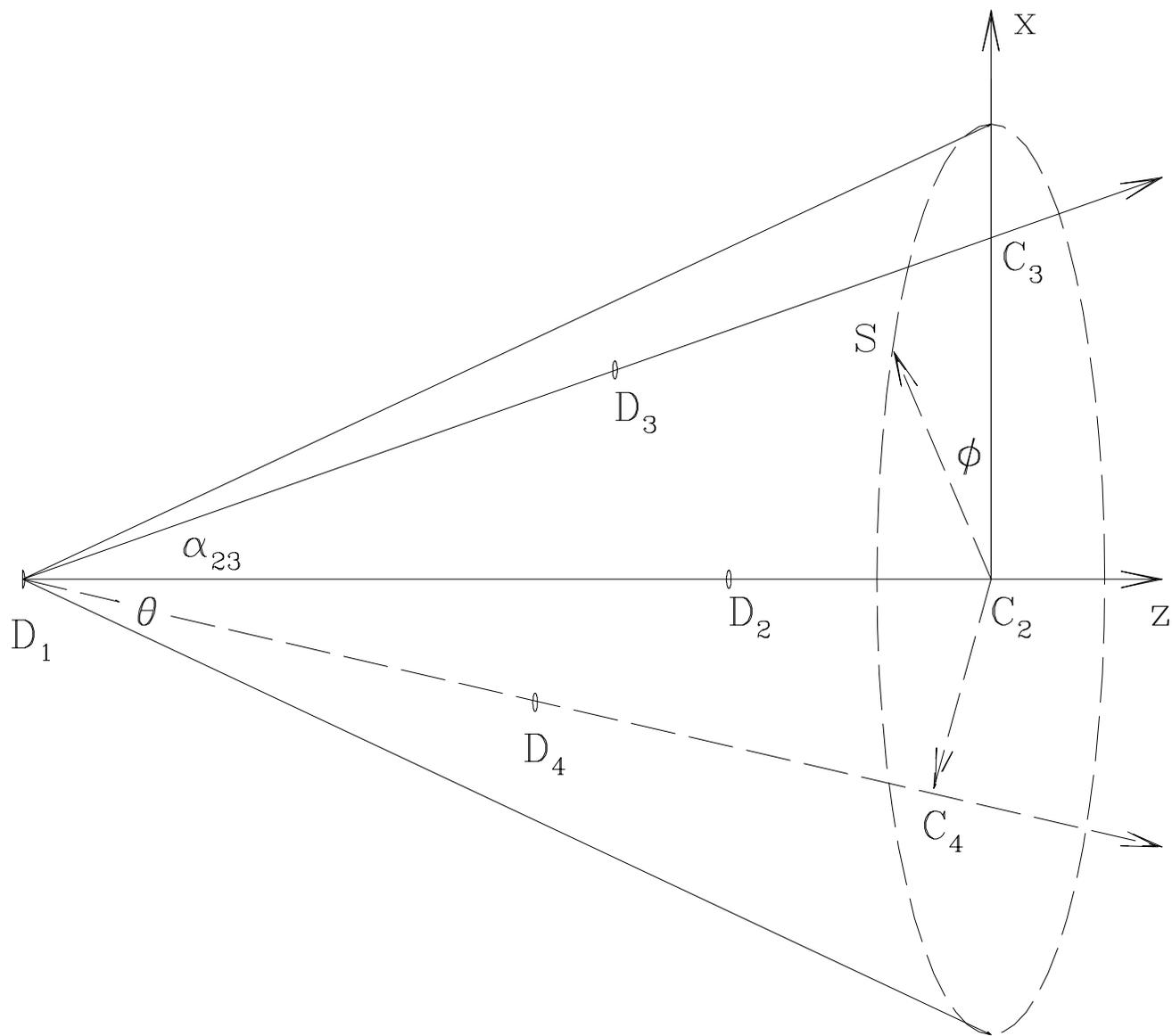
\begin{figure}
\caption{ The geometry of four detectors. Only solid lines and curves
represent those in the plane of the paper. $D_1$ - the reference
detector; $D_2$, $D_3$ - actual positions of detectors 2 and 3;
$\theta$, $\phi$ - direction of the source, S.}
\label{f1}
\end{figure}

As shown in Fig.\ref{f1}, let us take the line joining detectors $D_1$
and $D_2$ to be the $z$ axis. Let $n_2$ be the time delay, in units of
sampling interval $\Delta$, in the arrival times of the gravitational
wave between $D_1$ and  $D_2$ :
\begin{equation}n_2=(t_2-t_1)/\Delta,\qquad -N_2\le n_2 \le +N_2,
\label{n2}
\end{equation}
where $t_1$ and $t_2$ are arrival times at detectors $D_1$ and $D_2$
respectively; $\Delta$ represents the sampling interval. The quantity
$N_2$ [$=d_{12}/(c\Delta)$] represents the maximum time delay in terms
of measurements ($c \to$ the velocity of light, $d_{12}\to$ distance
between $D_1$ and $D_2$). The calculated values for the maximum
time-delay ($N_2$) between two of these detectors are given in Table
\ref{t00}.

Corresponding to each value of $n_2$, one can determine the polar
angle, $\th$ of the direction of wave:
\begin{equation}
\th=\cos^{-1}(n_2/N_2).
\label{th}\end{equation}
 One can now draw an imaginary circular disk, orthogonal to the $z$
axis, with center at any point $C_2$ on the axis, so that the line
joining $D_1$ and any point on the circumference of the circle makes
an angle, $\th$ with the $z$ axis, as shown in Fig.\ref{f1}.

Now we extend the line joining detectors $D_1$ and $D_3$, so that it
crosses the disk at point $C_3$. Let us take our $x$ axis to be the
line joining $C_2$ and $C_3$. If some point $S$ on the rim of the disk
represents the actual direction of the gravitational wave, then the
azimuthal angle, $\phi$, of this direction will be that between $x$
axis and the line joining points $C_2$ and $S$.

If $\beta_3$ is the angle between lines, ${\overline{D_1C_3}}$ and
${\overline{D_1S}}$, then the time-delay of the gravitational wave in
arriving $D_1$ with respect to $D_3$
can be given by $n_3=N_3\cos\beta_3,$ where $N_3$ is the maximum
time-delay between $D_1$ and $D_3$. The angle $\beta_3$ can be
expressed in terms of angles $\th$, $\phi$ and $\alpha_{23}$ which is
the angle between lines ${\overline {D_1C_2}}$ and ${\overline
{D_1C_3}}$. We thus obtain
\begin{equation}n_3 =
N_3\cos\th(\cos\alpha_{23}+\sin\alpha_{23}\cos\phi\tan\th).\label{n3}
\end{equation}

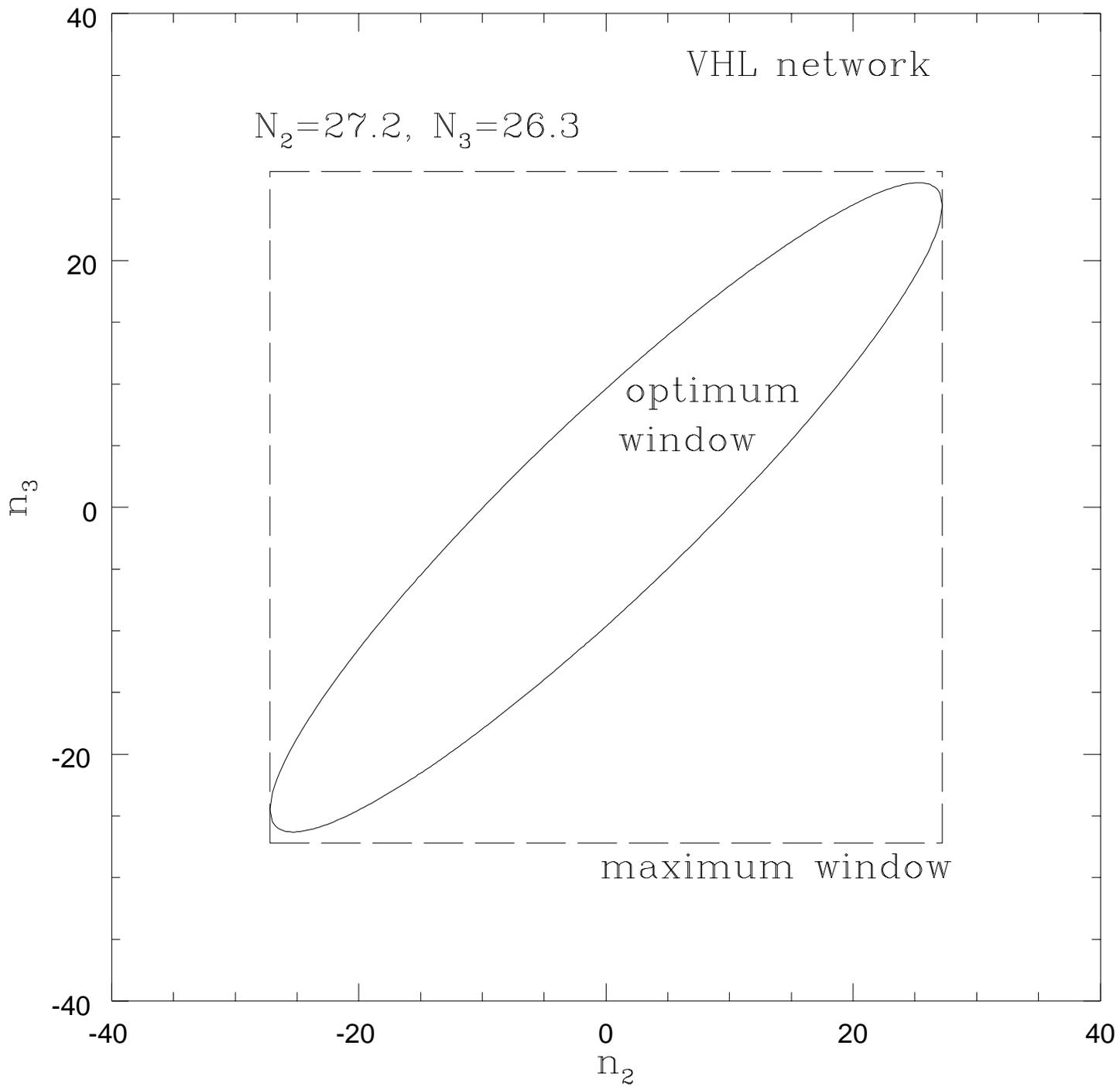
\begin{figure}
\caption{ V - VIRGO,  H - LIGO at Hanford, L - LIGO at Livingston. The
optimum window area in the shape of an ellipse for the HLV
network. The maximum window size is the square area enclosed by the
dashed lines.}
\label{f2}
\end{figure}

The maximum and minimum value of $n_3$ can be obtained when $\phi=0$
and $\pi$ respectively. In those cases, after some manipulation, we
arrive at
\begin{equation}n_3^2+n_2^2(p^2+q^2)-2pn_2n_3-q^2N_2^2=0,\label{n3ex}
\end{equation}

where $p=(N_3/N_2)\cos\alpha_{23}$ and $q=(N_3/N_2)\sin\alpha_{23}.$
This is the equation of an ellipse that thus represents the
circumference of the  surface of the required time delay window for
three detectors, as shown in Fig.\ref{f2} for the (H, L, V)
network. As shown in Fig.\ref{f3}, for any particular value of $n_2$
(say, 15), the value of $n_3$ may vary in between
$N_3\cos(\th+\alpha_{23})$ and $N_3\cos(\th-\alpha_{23})$, depending on $\phi$.

The window space is actually a lattice since the possible time delays
are integer multiples of the finite sampling time that we
consider. The area of this elliptical window is calculated to be
\begin{equation}{\rm Window\;size (3\;sites)}=\pi
N_2N_3\sin\alpha_{23}=2\pi\times {\rm Area\; of\; triangle\; joining\;
three\; detectors,}\label{win3}\end{equation}
which is an invariant quantity for any three detectors, irrespective
of the labelling of the detectors. The choice in labelling is,
however, important for the reason we discuss below.

We present values of the optimum window sizes for different networks
of any three  planned detectors in Table \ref{t1}.
It may be noted at this point that for the approximate calculation
presented in Ref.\cite{Shu}, the window size for three detectors was
overestimated to be the square of the largest value among the three
maximum time-delays expressed in number of measurements.
Our results show that the optimum window areas calculated in this way
can be, for example, as low as below 10\%   of their respective
maximum window areas for (H,V,G), (L,V,G), (V,G,A) and (V,G,T)
networks.

 We note that if a particular combination of $n_2$ and $n_3$ is known,
one can determine the azimuthal angle for the direction of
gravitational wave, $\phi$ to be any of these two possible values:
\begin{equation}\phi=\pm\cos^{-1}\Bigg[{N_2n_3-N_3n_2\cos\alpha_{23}\over
N_3\sqrt{N_2^2-n_2^2}\sin\alpha_{23}}\Bigg].\label{phi}\end{equation}
Along with Eq.\ref{th}, this provides two possible direction of the source.

\subsection{window size for four or more detectors}
 As mentioned in section \ref{one}, the four detector problem is over
determined. The calculation of window size depends on the procedure we
follow. If we measure the event amplitudes for four detectors then, in
principle, we need only one time delay to determine the solution that
involves five parameters and then we can reject noise-generated events
whose other time delays are not consistent with this solution. We,
however, would first like to check the consistency among the four time
delays and then measure the individual event amplitudes. This also
provides us an opportunity of testing Einstein's predictions regarding
the polarization states\cite{Gur}.

Let us now extend the line joining $D_1$ and detector $D_4$, so that
it crosses the imaginary circular disk at point $C_4$. Then the time
delay
of the gravitational wave in arriving $D_1$ with respect to $D_4$, is
$n_4 = N_4\cos\beta_4$, where $N_4$ is the maximum time delay between
$D_1$ and $D_4$
and $\beta_4$ is the angle between lines, ${\overline{D_1C_4}}$ and
${\overline{D_1S}}$. Expressing $\beta_4$ in terms of other angles, we
can write
\begin{equation}n_4=N_4[\cos\alpha_{24}\cos\th +
\sin\alpha_{24}\sin\th\cos(\psi-\phi)],\label{n4}\end{equation}
where $\psi$, the angle between the $x$ axis and the line joining
points $C_2$ and $C_4$, can be expressed as
\begin{equation}\psi=\cos^{-1}\Bigg[{\cos\alpha_{34}-\cos\alpha_{24}\cos\
alpha_{23}\over
\sin\alpha_{23}\sin\alpha_{24}}\Bigg].\label{psi}\end{equation}
Any angle $\alpha_{ij}$ ($i,j=2,3,4$) in above expresssions represents
the angle the line  joining detectors $D_i$ and $D_1$ make with that
joining $D_j$ and $D_1$.

So, corresponding to every set of values for ($n_2, n_3$), there can
be two possible values for $n_4$ depending on $\phi$ :
\begin{equation}n_4 = Xn_2+Yn_3\pm
Z\sqrt{K+Pn_2n_3-Qn_2^2-n_3^2},\label{n4b}\end{equation}
which has been obtained using expressions for $\th$ and $\phi$ in
terms of $n_2$ and $n_3$ and where the constants for a specific set of
detectors are represented by
\begin{eqnarray}
X& =&{N_4\over N_2}(\cos\a_{24}-\sin\a_{24}\cot\a_{23}\cos\psi),\nonumber\\
Y &= &{N_4\over N_3}\csc\a_{23}\sin\alpha_{24}\cos\psi, \nonumber\\
Z &= &{N_4\over N_3}\csc\a_{23}\sin\a_{24}\sin\psi, \label{fo}\\
K &= &N_3^2 \sin^2\a_{23}, \nonumber\\
P &= &{2N_3\over N_2}\cos\a_{23},\nonumber\\
Q &= &{N_3^2\over N_2^2}.\nonumber
\end{eqnarray}
One can now do some algebraic manipulation starting with squaring both
sides of Eq.(\ref{n4b}) to finally arrive at the equation of the
surface of an ellipsoid which is rotated with respect to its
coordinate axes. The two equations in Eq.(\ref{n4b}) with ($+$) and
($-$) sign in front of the square root represent two halves of the
ellipsoidal surface. Fig.\ref{f4} shows this surface for the
AHVL network with A as the reference detector.

The events that may lead to the false alarm will be points distributed
on such a surface. So, once we choose the reference 3-detector
subnetwork, the window size for the 4-detector network would be the
number of lattice cells on the ellipsoid which is twice the window
size for the first three detectors.:
\begin{equation}{\rm Window\;size (4\;sites)}=2\pi
N_2N_3\sin\alpha_{23}=4\pi\times {\rm Area\; of\; triangle\; joining\;
3\; reference\; detectors,}\label{win4}\end{equation}
 We observe that the maximum window size of a three dimensional volume
($=W^3$), where $W/2$ is the maximum of the values of time delay among
the detectors gets reduced to the area of a two dimensional surface in
three dimensions. We can, therefore, expect that the optimization
would  help to a substantial extent in reducing the threshold level
calculated otherwise using maximum window area.

Now we address the question how one should select the reference
subnetwork of three detectors out of four possible choices. We can
have two criteria of selection: (i) The window area of the reference
subnetwork should be as low as possible to avoid increase in the false
alarm rate; (ii) the unavoidable presence of uncertainty in
measurements of the time delays will always allow some room in the
window for noise-generated events. So, the reference subnetwork should
be chosen in such a way that the error in  determining the direction
of the source from its time-delay data is as low as possible.

{}From Eqs.(\ref{th}) and (\ref{phi}), we note that the accuracy of the
determination of direction improves with larger values of $N_2$, $N_3$
and $\alpha_{23}$ for  given values of error, $\Delta n_2$ and $\Delta
n_3$. So, it is always better to choose the reference detector in such
a way that these values take the largest possible ones. One has to
make a good choice by considering all three possible sets of values.
These selection criteria for achieving smaller errors in the
determination of  the direction of source, however, go against a
choice based on the smallest window size: The accuracy is the best for
the 3-detector subnetwork with the largest window area.

Since these two criteria make diametrically opposite demands on the
window size, we would consider their quantitative effects. Referring
to Table \ref{t1}, for example, one can see that for the AHVL network,
the highest possible window size corresponding to the reference
subnetwork, AHV is about 3.7 times the lowest possible window size
corresponding to the reference subnetwork, VHL.
Since the threshold would be normally at quite high level (say,
$7\sigma$ or even higher; see next section for more discussion and the
definition of $\sigma$, the standard deviation of noise), increasing
the window size by a factor of 3 or 4 has only a small effect on the
thresholds, but it could lead to dramatic improvement in the accuracy
for determining the direction of the source.

We, therefore, ignore the criterion (i) and follow only the criterion
(ii) for making our selection for the reference detector and the
reference subnetwork of three detectors.
So, for example, in case AIGO is a member of the four detector
network, we would always like to choose it as the reference one (the
$D_1$) simply because it is always at the farthest distance from any
of the detectors. Similarly, we would also choose H to be another
member of the reference subnetwork because this provides another
half-length detector at the same site and thus extra advantage in
coincidence detection. So, as an example, for (AHVL) network, our
choice of the reference subnetwork is (AHV). We choose V instead of L
because the angle $\alpha_{23}$ is the largest in this case. From now
onwards, whenever we refer to a network of four or more detectors, the
first three names written will represent our choice of the reference
subnetwork and the first name will correspond to the reference
detector.

Once one does observation with a four detector network one determines
the direction of the source and thus if a fifth detector is available,
one just has to check whether or not the time of arrival in that is
consistent with the values of the four-detector network. So, we can
conclude that the optimum window size for a network with detectors at
five or more different sites is again just twice the size of the
window for the reference 3-detector network.

\section{Setting Thresholds for detectors}
\label{4th}

We assume that in the output $x$ of any of the detectors, $\j$,  the
only noise source is stationary and characterised by the Gaussian
Normal distribution, with zero mean and standard deviation,
$\sigma_\j$. The values of $\sigma_\j$ are, of course, different for
different detectors. So, the probability that the output $x$ will
exceed some fixed threshold value, $X_0$ is given by
\begin{eqnarray}
P(\mid x\mid >X_0) &=& 1-2\int\limits_0^{X_0}\,{1\over\sqrt{2\pi}\sigma_\j}
                    \exp[-x^2/(2\sigma_\j^2)]\, dx\nonumber\\
                  &=& {\rm erfc}({\cal T}_\j/\sqrt{2}),
\end{eqnarray}
where
\begin{equation}
{\cal T}_\j = {X_0\over \sigma_\j}
\end{equation}
which, from, now onwards will be referred to as the value of the {\it
threshold} set in the detector $\j$. The $\sigma_\j$ here is the value
of the standard deviation of the noise averaged over the bandwidth of
the detector. For the white noise case, for example, $(\sigma_\j)^2$
is just the bandwidth divided by $F^\j_{\rm bur}$.

It is advantageous to choose a low threshold, and correspondingly high
false alarm rate for the less sensitive detectors. The noise-generated
accidental coincidence rate is kept down mainly by the lower rate of
false alarms in the more sensitive detectors. So, we keep the value of
the output $X_0$ corresponding to the threshold to be same in all the
detectors. As pointed out in section \ref{third}, this assumption
ignores the fact that different detectors will actually give different
responses to a gravitational wave depending on their orientations on
the surface of the earth and the direction and  polarization angle of
the wave.

So, if $X_0$ is kept at the same level in different detectors by
neglecting the above-mentioned effect, the values of ${\cal T}_\j$ is
to be set at different values in different detectors.
If ${\cal T}_1$ is the threshold of the first detector, then, in order
to detect the same gravitational wave, one should set the threshold of
detector $\j$ at $\zeta^\j {\cal T}_1$, where $\zeta^\j$ is the ratio
of the SNRs of the $\j$-th and the first detector.

So, on the basis of our discussion in the last three sections, we can
now fix the thresholds, ${\cal T}_\j$ in a network having $n$ number
of detectors by the following equation:
\begin{equation}
{\rm erfc}({\cal T}_1/\sqrt{2})\prod\limits_{\j=2}^{n} {\rm
erfc}(\zeta^\j{\cal T}_1/\sqrt{2})={{\cal F}\over W \chi}
\label{thres}
\end{equation}
where W is the window size and $\chi$ is a factor equal to unity for a
burst source and equal to the number of filters, $N_f$ for a
coalescing binary signal. We consider a false alarm rate, ${\cal F}$,
of one per year or $3.171\times 10^{-11}$ in our calculation.  As our
discussion in the following subsections will show, for a single
detector observing coalescing binaries, this translates into a
threshold value of 7.65, whereas for the same observing millisecond
burst sources, the threshold has to be set at 6.64 due to the absence
of filtering in the latter case. Also for the purpose of setting
setting the threshold, we ignore the correlation between the samples
of the statistics used for detection. Although there is some
correlation among the samples, the effect on the threshold has been,
however, shown to be negligible\cite{ShDh}.

We, therefore, decide, that a detector whose sensitivity is less than
about one-third of any other detector in the same network does not
come to be very useful for the network as such. In that case the false
alarm rate of the less sensitive detector would be very high, or
equivalently, the noise generated events in that detector would cross
the threshold quite often and such a coincidence experiment would not
carry any meaning at all. For instance, in case of two detectors, such
an experiment would be almost nothing different from the more
sensitive detector observing alone.

One can, therefore, guess from these values of $\zeta^\j_{\rm cb}$ in
Table \ref{t20} that the medium scale
 interferometers (G, A, T) under power recycling operation will not be
very useful for a coincidence experiment with the LIGO-VIRGO
network. However, as discussed in section \ref{second}, if
 they are narrow-banded by dual recycling, say, with a bandwidth of
300Hz around
 250 Hz as planned for GEO\cite{Win},  their sensitivities become
comparable, and in some cases, better than that of LIGO or VIRGO with
power recycling operation. A meaningful coincidence experiment can,
therefore, be performed only under this condition. For the same
reason, TAMA, even with its length of 300m, may come to be very useful
for the networks of the rest of the planned detectors,  especially for
obtaining a good sky coverage and accuracy in the determination of the
source parameters, once an improved pendulum is incorporated in the
design.

\subsection{Thresholds for detecting signals from burst sources}

We should recall here that in the absence of our knowledge about the
possible waveform of the burst sources, we assumed that the power of
burst is uniformly distributed over a frequency range
(70-2000)Hz. Setting thresholds for detecting such burst sources is
relatively easier as compared
to the case of coalescing binary signals due to the absence of
filtering. The millisecond burst sources will be detected in
coincidence experiments just by  checking consistency in the arrival
of burst in the time series of different detectors. In this case, we
set $\chi=1$ in the right hand side of Eq.(\ref{thres}). So, if we
consider a single detector observing millisecond burst sources, for a
false alarm rate, ${\cal F}$ of one per year, the threshold, ${\cal
T}$  has to be set at 6.64.

For the model of burst waveform
considered by us, we find that the relative sensitivity of the
power-recycled VIRGO is quite high as compared to others with their
present design parameters and thus, as discussed above, a  coincidence
experiment  performed by including VIRGO in the network would not be a
meaningful one.

In Table \ref{t52} we present the threshold values we have set for
different detectors participating in the coincidence experiments of
some interesting networks for the observation of  millisecond burst
sources.

 We find  that a network of power-recycled LIGOs  and dual-recycled
GEO and AIGO can increase the volume of the sky covered by about 4.9
times as compared with only power-recycled LIGO detectors for the
observation of the burst sources of a given strength. The absolute
range of observation will, of course, vary with the strength of the
burst.

\subsection{Thresholds for detecting signals from coalescing compact binaries}
\label{4B}

The choice of the threshold levels for the detection of signals from
coalescing compact binaries will differ from that for burst sources
due to the effect of filtering. Filtering can have the following three
effects on the setting of thresholds:
\begin{description}
\item{(i)} If it is assumed that there is no correlation in the filter
outputs, then $N_f$ number of filters lead to an increased false alarm
rate of $(N_f{\cal F}_0)$, where ${\cal F}_0$ is the false alarm rate
when no filter is used. Here we set $\chi=N_f$ in Eq.\ref{thres}.

The problem of determination of the lattice of filters for the
observation of coalescing binary signals was investigated by
Sathyaprakash and Dhurandhar\cite{BSS} for the white noise case and
again by Dhurandhar and Sathyaprakash\cite{SVD} for the coloured noise
case. But, as discussed below, in the present case the problem is
somewhat nontrivial, although not very important for setting threshold
levels of various detectors.

In order to apply the technique of matched filtering for the detection
of the chirp waveform, it is necessary to construct a lattice of
filters corresponding to different values of the three parameters in
their relevant range. The spacing between filters depends on the
allowed value of drop in the correlation function of two {\it chirp}
waveforms with different values of their parameters.
It was shown\cite{BSS} that for the Newtonian chirp waveform, it is
sufficient to choose such a set of filters covering only the range of
the phase $\phi_c$ and the chirp mass, ${\cal M}$. However, since the
correlation of a given data set with a filter of arbitrary phase can
always be expressed as a linear combination of the correlation of the
same with two independent basis filters corresponding to, say,
$\phi_c=0$ and $\phi_c=\pi/2$, it was shown\cite{BSS} to be sufficient
to construct a lattice with two filters for each value of the mass
parameter. Once the correlations are calculated with the basis sets,
the phase of the filter which maximises the correlation can be easily
calculated by an analytical expression.

However, the spacing between mass parameters for the choice of filters
crucially depends on the  noise characteristics of the detector.  For
example, it was found\cite{SVD} that the consideration of coloured
noise in the case of power recycling led to the reduction of the
number of filters by approximately half as compared to that
considering white noise. This reduction in number is mainly due to the
effective  reduction of the bandwidth which happens because, in case
of recycling, the sensitivity of a detector is enhanced at the lower
end of the detector bandwidth at the cost of a greater noise at higher
frequencies.  This leads to a slow fall-off in correlation and thus to
larger spacing between filters.

Since the noise characteristics of different detectors would be
different depending on their choice of the knee frequency, laser
power, laser wavelength, mirror loss etc., so, strictly speaking, the
distribution as well as the number of filters in the lattice can be
chosen to be different for different detectors. So, in the present
context, we have to define properly what we would call a `coincidence
detection' by analysing the outputs of various detectors in a
network.

For a coincidence detection, it is necessary that the filters with the
same set of values of the parameters of the chirp waveform in
different detectors cross
their thresholds at appropriate times. Since the phase, $\phi_c$
observed by a detector is dependent on its orientation, we neglect the
matching in $\phi_c$ as a criterion for the coincidence detection.
So, we consider an event to be a case of `coincidence detection' if
filters with the same value of the mass parameter in different
detectors cross their thresholds
at appropriate points in time. The information obtained about $\phi_c$
can be important, however, for a more rigorous second analysis of the
data once a decision is arrived on coincidence detection based on the
above criterion.

On the other hand, if the same range of the mass parameter is covered
by different sets of filters, $S_F^\j$, characterised by differently
selected values for the mass parameter, then it may so happen that
only one detector detects a signal while others miss it due to the
absence of the filter corresponding to that particular one in their
lattice. But if such a situation arises, the other detectors can then
do a reanalysis, searching for the corresponding signal this
time. This shows that the set of filters that will actually contribute
to the false alarm rate is the union of these different sets for
different detectors, i.e., $\bigcup\limits\j S_F^\j$.

A detailed analysis leading to the accurate determination of the
number of filters for different detectors is, however, unimportant for
the main objective of this paper. Our interest in the number of
filters is due to  its effect on setting thresholds as mentioned
above. We can see that although the order of magnitude of the number
obtained by multiplying the window size with $N_f$ is very important
for setting the thresholds, a factor of 2 or 3 in the ratio of the
number of filters for different detectors does not make much
difference in the threshold levels.
We, therefore, assume that, to avoid complications, the same number of
filters, $N_f$ will be chosen for different detectors participating in
the coincident experiment, irrespective of their noise
characteristics, thus demanding same computational facility in all the
detectors. For example, the number of filters calculated for the most
sensitive detector, but with the highest value of the sampling rate
(corresponding to the widest bandwidth one likes to consider) may be a
good choice for this common number.

For the present analysis, however, we take a conservative estimate of
$N_f$ (a maximum value) which can be obtained {\it by considering the
white noise case}.         The number of filters obviously depends on
the range of chirp mass one intends to cover and for a high value of
the upper limit (say, ${\cal M}=20 M_\odot$), $N_f$ for the white
noise case can be given by Eq.(4.2) of Ref.\cite{BSS}:
\begin{equation}
 N_f = 3000\Bigg[{{\cal M}_1\over M_\odot}\Bigg]^{-5/3}
\Bigg[{f_l\over 100 {\rm Hz} }\Bigg]^{-8/3} \Bigg[{\Delta \xi\over
1{\rm msec}}\Bigg]^{-1},
\label{NFilter}
\end{equation}
where ${\cal M}_1$ is the lower limit on the chirp mass and  $\Delta
\xi$ is the spacing between the filters corresponding to the values of
the chirp mass determined by the following relation between ${\cal M}$
and $\xi$ which is referred to as the {\it sweep time} of the chirp
waveform:
\begin{equation}
\xi = 3.00\Bigg[{{\cal M}\over M_\odot}\Bigg]^{-5/3}\Bigg[{f_l\over
100{\rm Hz}}\Bigg]^{-8/3} {\rm sec.}
\end{equation}
The spacing, $\Delta \xi$ is almost a constant for a particular chosen
value of the allowable drop in the correlation function between two
neighbouring filters. We note that our upper limit of frequency, $f_u$
corresponds to a total mass of the binary, $M=2.2 M_\odot$ and thus a
lower limit on the chirp mass, ${\cal M}_1 = 1.0 M_\odot$ is a
suitable choice to make. For an allowable drop in correlation to 0.90
between neighbouring filters, the number of filters is calculated to
be approximately 1600. This is the value that we use in our
calculation of the thresholds.

It should be noted that the Eq.(\ref{NFilter}) has been evoked here
just to get an estimate of the number of filters that may be necessary
to detect the coalescing binary signals. However, it should not be
taken literally for the actual detectors, since their noise profile
differs significantly from the noise modelled to arrive at this
equation: white noise for $f\ge f_l$ and a steep wall below $f_l$.

\item{(ii)} On the other hand, if data are filtered, that also reduces
the effective noise bandwidth, so the rate at which it needs to be
sampled to give an accurate picture of it is less. Since the
false-alarm probability depends on how many randomly chosen samples
one has, therefore, the rate at which false alarms occur will be less
\cite{Shu}.

\item{(iii)} The auto-correlation function of the filtered output
resembles the sinc function and so, contrary to the assumption made in
item(i) above, the adjacent sample points are correlated and,
therefore, the filtered output is actually coloured.

These two problems, (ii) and (iii), described above have been studied
in detail by Dhurandhar and Schutz \cite{ShDh} and the conclusion is
that the reduction of the false alarm rate due to filtering is not
much. We can safely ignore these effects in our calculation of the
threshold levels.

\item{(iv)} The error induced by filtering in exact determination of
the time of arrival or time of coalescence of the binary:  Any error
in determination of these times will lead to a probabilistic region
around every lattice point of the window surface and may thus
seriously affect our calculation of the window size described in
sec. \ref{third}.

The detailed calculation \cite{BSD} shows that whereas this error is
of the order of $\pm$20-25msec (with 68\% confidence) for the time of
arrival, the same figure is of the order of $\pm$0.4-0.6 for the time
of coalescence for Newtonian and post Newtonian cases. Therefore, it
seems to be better to choose the time of coalescence rather than the
time of arrival in our window size calculation. In that case the
existence of the probabilistic region around the exactly determined
window will not contribute much correction to this.

\end{description}

In Table \ref{t51} we present the threshold values we have set for
different detectors participating in the coincidence experiments of
some interesting networks for the observation of signals from the
coalescing compact binary sources.

We can also calculate the range of observation by these networks by
using Eq.(5.8) of Finn and Chernoff\cite{FCher} which we translate to
our case as
\begin{equation}
{\cal R}_{90\%}=\Bigg({{\cal M}\over
1.2M_\odot}\Bigg)^{5/6}\Bigg({7.65\over{\cal T}_L}\Bigg)\Bigg({F_{\rm
cb}^\j\over 5.0\times 10^{42}}\Bigg)^{1/2}\times 37.67 {\rm Mpc}.
\end{equation}
 One should note that not all sources within a particular range,
${\cal R}$ are detectable and, on the other hand, some sources
outside ${\cal R}$ may come to be detectable. So, ${\cal R}_{90\%}$ is
defined as the distance within which 90\% of the observable sources
would lie. The chirp mass, ${\cal M} =1.2M_\odot$ corresponds to,
among other combinations, a 1.4-1.4 $M_\odot$ system like the
Hulse-Taylor binary pulsar, PSR1913+16.

This leads to a detection rate of compact coalescing binary sources
(see Eq.(5.5) of Finn and Chernoff\cite{FCher}) to be
\begin{equation}
\Bigg({{\dot {\cal N}}\over 8\times 10^{-8} {\rm Mpc}^{-3}\,{\rm
yr}^{-1}}\Bigg)
\Bigg({{\cal M}\over 1.2 M_\odot}\Bigg)^{5/2}\Bigg({7.65\over{\cal
T}_L}\Bigg)^3
\Bigg({F_{\rm cb}^\j\over 5.0\times 10^{42}}\Bigg)^{3/2}\times
4.26\times 10^{-3}\;{\rm yr}^{-1}
\end{equation}
The expectation value for the number density of sources per unit time,
${\dot {\cal N}}$  was estimated by Phinney\cite{Phin},
which ranges from an ultraconservative value of \{$6\times 10^{-10}$\}
Mpc$^{-3}$ yr$^{-1}$ to a conservative estimate of \{$8\times
10^{-8}$\} Mpc$^{-3}$ yr$^{-1}$ and to an upper limit of \{$6\times
10^{-5}$\} Mpc$^{-3}$ yr$^{-1}$.

We present the values of the range and the rate of detection of
coalescing binaries for some interesting networks in Table
\ref{t53}. We note that a coincidence experiment of power-recycled
LIGOs and VIRGO with dual-recycled GEO and AIGO can increase the
volume of sky covered by only LIGO detectors  by 3.2 times and the
same covered by the LIGO-VIRGO network by about 1.7 times. This is of
course far less than the range that can be covered by LIGOs and VIRGO
with dual recycling operation at a later stage, but the latter case is
not very useful for the determination of the direction, distance and
other source parameters, as described in sections \ref{one} and
\ref{second}.

\section{concluding remarks}
\label{6th}
In this paper we investigated how effectively the planned initial
interferometeric gravitational wave detectors would be able to detect
compact coalescing binaries and burst events and what volume of the
sky can be covered by them in the beginning of the next century. Here
we summarize our main results:

\begin{description}
\item{(i)} We made a comparison of the relative sensitivities of all
the planned detectors for the observation of both of the
above-mentioned sources by taking into account all possible kinds of
noise sources which would affect their performance. The numbers
obtained by us for the relative SNRs may, however, vary because every
detector leaves room for incorporating better techniques to reduce the
thermal and the seismic noise. But to a very good approximation these
numbers may be taken to be the representative ones for the
sensitivities of the detectors in their initial stages. We also
calculated relative SNRs for LIGO, VIRGO and TAMA with dual recycling
configuration, although they do not have any plan of implementing this
in the first stage of their initial detectors.

\item{(ii)} From the values of the relative SNRs of different
detectors, we find that a meaningful coincidence experiment can be
performed if the detectors have similar sensitivities. Thus, if LIGOs
and VIRGO are operated in the power recycling mode and all other
medium-scale detectors are operated in dual recycling mode, the
sensitivities for the detection of the coalescing binary signals
become comparable and a meaningful coincidence experiment can be
carried out. Similarly, we find that the sensitivity of even the
power-recycled VIRGO for the detection of the type of burst sources
considered by us (i.e. uniform power spectral density upto 2000 Hz) is
quite high as compared to other detectors in their initial stage. So,
although VIRGO alone will be able to cover a large volume of sky for
the detection of this type of burst sources, it will not come to be
useful for a coincidence experiment together with  other detectors in
a network.

\item{(iii)} Next we calculated the time-delay window sizes for
networks with all possible combination of detectors. We also set a
criterion whereby one should select a reference detector as well as a
reference combination of three detectors for the determination of the
direction of the source. This is necessary for effectively reducing
the error in the determination of the direction of the source: The
three reference detectors chosen should form a triangle with maximum
area. For example, AIGO should always be chosen as the reference
detector since it is always at the farthest distance from any other
detector. Similarly, LIGO at Hanford should also be chosen as another
member of the reference 3-detector subnetwork because it provides one
more detector of 2000m length at the same site and thus extra
advantage in coincident experiment.

\item{(iv)} We discussed in detail the relevant issues of filtering to
calculate the threshold levels for the detection of the coalescing
compact binaries. The important issue of how comparison should be made
among the outputs of various detectors using different lattices of
filters covering the same range of source parameters was also
discussed. We finally made a conservative estimate (a maximum number)
of the number of filters which would affect the rate of false alarm
for the coincidence experiment.

\item{(v)} {\it Threshold criteria for burst sources and coalescing binaries:}
We have defined relative sensitivities, $\zeta^\j$ which depend on the
type of source and the detector $\j$ and in terms of these quantities
have set up the threshold levels for two types of sources, namely, (a)
burst sources, and (b) coalescing binaries.

(a) In the absence of our knowledge of the possible waveform of the
burst sources, we assumed that the power of burst is uniformly
distributed over a frequency range upto  2000 Hz. We find, by setting
thresholds for different detectors,  that a network of power-recycled
LIGOs  and dual-recycled GEO and AIGO can increase the volume of the
sky covered by only power-recycled LIGO detectors  by about 4.9 times
for the observation of the burst sources of the same strength. The
absolute range of observation will vary with the strength of the
burst.

(b) We find that a coincidence experiment of power-recycled LIGO
detectors and VIRGO and dual-recycled GEO and AIGO can increase the
volume of the sky covered by 3.2 times as compared with only
power-recycled LIGO detectors and by 1.7 times as compared with the
power-recycled LIGO-VIRGO network.
This is of course far less than the range that can be covered by only LIGOs or
the LIGO-VIRGO network with dual recycling operation, which they may
implement in the later stage of their `initial' interferometers but
the accuracy in the determination of direction, distance and other
source parameters will be much better in a coincidence experiment in
which other detectors and especially AIGO can take part.

\end{description}

We suggest further investigations which need to be carried out on
various issues related to this paper. As discussed in item (iv) above,
although the actual number of filters may vary for different
detectors, we made a conservative estimate of the number of filters to
calculate the threshold levels. A detailed analysis, however, is in
order taking into account several filter banks corresponding to
different detectors. Secondly, the phase information can be used along
with the time-delay information to decide on coincident detection. For
this purpose the orientations of the detectors will have to be taken
into account. As discussed in section \ref{4B}, it is also important
to carry out a detailed investigation for the determination of the
direction of the binary and thus also the window size using
time-delays in the {\it time of coalescence} (instead of {\it time of
arrival}) between different detectors (also see Ref.\cite{BSD}). On
similar lines as above, we could also treat other sources which have
different power spectra. An important source that may be mentioned in
this connection is the stochastic gravitational radiation. An analysis
applying the results of the detailed investigations carried out by
Christensen\cite{Chris} and Flanagan\cite{FLA} to the planned
interferometers may come to be useful for the evaluation of the
sensitivities of the networks for the detection of the stochastic
radiation.

\acknowledgements

We would like to specially thank J. Sandeman for suggesting to us this
problem and  D. Shoemaker for clarifying various points and for giving
valuable comments on a preliminary version of this manuscript. It is
also our pleasure to thank D. Blair, M. Fujimoto, D.E. McClelland,
M. Taniwaki, J.-Y. Vinet and W. Winkler for clarifying some important
points and supplying various information and R. Balasubramanium,
S.D. Mohanty and B.S. Sathyaprakash for useful discussions.

\begin{figure}
\caption{ V - VIRGO,  H - LIGO at Hanford, L - LIGO at Livingston.
The time-delay between V and L, $n_3$ as a function of the angle
$\phi$ of the direction of source for a fixed value (=15.) of the time
delay, $n_2$ between V and H. }
\label{f3}
\end{figure}

\begin{figure}
\caption{The ellipsoidal surface for the time-delay, $n_4$ of the AHVL network.
A - AIGO, V - VIRGO,  H - LIGO at Hanford, L - LIGO at Livingston.}
\label{f4}
\end{figure}

\begin{table}
\caption{ Description of the site and configuration of the planned
detectors. The letter written in the second column represents the
corresponding detector throughout our analysis. The orientation refers
to the angle made by the bisector of the arms with the local south
$\to$ north direction measured in anticlockwise convention. DL and FP
in the 4th column refers to delay lines and Fabry-Perot cavities
respectively, which are incorporated in the arms. The recycling
configuration indicated in the fifth column is for the initial stages
of the interferometers. All these detectors will have provision for
incorporating dual recycling at later stages. The minimum length of
the proposed AIGO detector has been set to be 500m, however, this may
change, but not much, depending on the budget available. }
\label{t0}
\begin{tabular}{lccccccr}
Detector & Denoted by & Length & Arms & Recycling & Latitude &
Longitude & Orientation\\ \hline
 LIGO, Hanford, Washington & H4 & 4000m & FP & Power &
46$^0$27$^\p$18.5$^{\pp}$N & 119$^0$24$^\p$27.1$^{\pp}$W & 81.8$^0$ \\

 LIGO, Hanford, Washington & H2 & 2000m &FP & Power &
46$^0$27$^\p$18.5$^\pp$N & 119$^0$24$^\p$27.1$^\pp$W & 81.8$^0$\\
 LIGO, Livingston, Lousiana & L & 4000m & FP & Power &
30$^0$33$^\p$46.0$^\pp$N & 90$^0$46$^\p$27.3$^\pp$W & 153$^0$\\
 VIRGO, Pisa, Italy & V & 3000m & FP & Power & 43.3$^0$N & 10.1$^0$E &
26.5$^0$\\
GEO, Hannover, Germany & G & 600m & DL(3 bounces) & Dual &
52$^0$14.8$^\p$N & 9$^0$49.3$^\p$E& 339.25$^0$ \\
AIGO, Perth, Australia & A & $\sim$ 500m & DL(3 bounces) & Dual &
31.04$^0$S & 115.49$^0$E & {\sl not decided} \\
TAMA, Mitaka, Japan & T & 300m & FP & Power & 35$^0$40$^\p$25$^{\pp}$N &
139$^0$32$^\p$15$^{\pp}$E &  225$^0$  \\
\end{tabular}
\end{table}

\begin{table}
\caption{Values of the parameters used in the calculation of the
relative signal to noise ratio of the detectors. In most of the cases,
these have been obtained by personal communication with people
involved in designing the planned detectors. Some of these are quoted
as {\it conservative estimates}, while others are quoted to be {\it
hopefully achievable ones}. Values marked with (*) are assumed by
us. The value of $R_1$ for H2 has been chosen such that its cavity
finesse becomes approximately twice that of H4. It should be noted
that
the TAMA (300m) interferometer has been designed to work as a
prototype in its initial stage.}
\label{t10}
\begin{tabular}{lcccccc}
 Quantities    &L,H4  &H2     & V            & G        & A   & T\\ \hline
End mass, $m$(Kg)  &10 &10 & 42 &16  &16   &1\\
Suspension quality, $Q_0$ &$10^7$&$10^7$ &$3\t 10^5$ &$5\t 10^7$&
$5\t 10^7$&$10^5$\\
End mass fundamental mode, $f_{\rm int}$(KHz)   &23    &23 &5.7
&20$^*$          &20$^*$     &30\\
End mass quality, $Q_{\rm int}$ &$10^5$&$10^5$&$10^6$&$5\t
10^6$&$5\times 10^6$     &$5\t 10^5$\\
Laser power, $\eta I_0$(watts) &5  &5 &10  &5  &5    &3 \\
Laser wavelength, $\lambda$($\mu$meter)&0.514 &0.514   &1.06 &1.06 &1.06&1.06\\
Mirror losses, $A^2$($\t 10^{-6}$)  &50.   &50.&2. &20.    & 20. &10. \\
Input mirror reflectivily of cavity,$R_1$ &0.97&0.985$^*$&0.85 & -&-&0.988\\
\end{tabular}
\end{table}

\begin{table}
\caption{Relative SNRs of the detectors, $\zeta^\j_{\rm cb}$ and
$\zeta^\j_{\rm bur}$  as normalised with respect to that of full
length LIGO using power recycling {\it for the same kind of source}
. A - AIGO, H4 - 4000m LIGO at Hanford, H2 - 2000m LIGO at Hanford, L
- LIGO at Livingston, G - GEO, V - VIRGO, T - TAMA. The dual recycling
parameters are, in all cases, chosen to be those for a bandwidth of
300Hz around 250Hz as will be used in GEO. It should be noted that for
different detectors, maximum gain may be obtained for some different
values for bandwidth and central frequency. $^\dagger$Although the
dual recycling values for LIGO, VIRGO and TAMA have been calculated,
but they do not have any plan to implement this at the initial stage
of their interferometers. It should also be noted that the TAMA (300m)
interferometer has been designed to work as a prototype in its initial
stage.}
\label{t20}
\begin{tabular}{lcccc}
Detector &$\zeta^\j_{\rm cb}$(Power)& $\zeta^\j_{\rm cb}$(Dual)&
$\zeta^\j_{\rm bur}$(Power)    &$\zeta^\j_{\rm bur}$(Dual)    \\
\hline

L, H4 &1.0  & 2.78$^\dagger$  &1.0  & 6.03$^\dagger$   \\
H2    &0.58 & 1.38$^\dagger$  &0.66 & 3.01$^\dagger$    \\
V     &1.25 & 1.93$^\dagger$ &4.33 & 9.36$^\dagger$ \\
G     &0.13 &0.95   &0.15 & 1.82     \\
A     &0.11 &0.79   &0.13 & 1.51     \\
T     &0.028&0.034$^\dagger$       &0.168     &0.230$^\dagger$   \\
\end{tabular}
\end{table}

\begin{table}
\caption{ Maximum time delays, $N_2$ between two detectors. A - AIGO,
H -  LIGO at Hanford, L - LIGO at Livingston, G - GEO, V - VIRGO, T -
TAMA.}
\label{t00}
\begin{tabular}{lclclc}
Detectors & Maximum delay & Detectors & Maximum delay& Detectors &
Maximum delay\\ \hline
HL &10.0 & HV & 27.2 & LV & 26.3 \\
HG &25.0 & LG & 25.0 & VG & 3.3  \\
HA &39.2 & LA & 41.6 & VA & 37.0 \\
GA &37.3 & HT & 14.6 & LT & 22.0 \\
VT &21.4 & AT & 35.2 & GT & 18.5 \\
\end{tabular}
\end{table}

\begin{table}
\caption{Window sizes for 3-detector networks. A - AIGO, H - LIGO at
Hanford, L - LIGO at Livingston, G - GEO, V - VIRGO, T - TAMA. }
\label{t1}
\begin{tabular}{lclc}
Network & Optimized area& Network & Optimized area\\ \hline
VHL& 822.2& GHL& 769.5\\
AHL& 1222.6& LVG& 243.9\\
ALV& 3012.8& HVG& 201.3\\
AHV& 3031.3& ALG& 2894.9\\
AHG& 2831.0& AVG& 383.2\\
THL& 372.0& THV& 975.6\\
THG& 841.4& TGV& 98.4\\
TLG& 1238.9& TLV& 1425.9\\
ATL& 2432.8& ATV& 2309.7\\
ATG& 2023.9& ATH& 1609.1\\
\end{tabular}
\end{table}

\begin{table}
\caption{Threshold to be set for different detectors for coincident
observation of millisecond burst sources. A - AIGO, H4 - 4000m LIGO at
Hanford, H2 - 2000m LIGO at Hanford, L - LIGO at Livingston, G - GEO,
V - VIRGO. H - represents both detectors at Hanford together. For
networks with detectors at more than three sites, the first detector
represents the reference detector, whereas the first three detectors
represent the reference 3-detector network, on the basis of which we
calculated the window size ({\it see text}). The superscripts (\b) and
(\n) represent the power and dual recycling modes of opration
respectively.}
\label{t52}
\begin{tabular}{lcccccc}
Network of detectors               & H4   & H2   & L    & V    & G    & A   \\
\hline
H4\b H2\b                          & 5.37 & 3.54 & -    & -    & -    &-    \\
H4\b H2\n                          & 2.05 & 6.17 & -    & -    & -    & -   \\
H4\b L\b or H4\n L\n               & 4.85 & -    & 4.85 &-     & -    & -   \\
H4\b H2\b L\b                      & 4.30 & 2.84 & 4.30 & -    & -    & -   \\
G\n H4\b H2\b L\b                  & 2.95 & 1.95 & 2.95 & -    & 5.37 & -   \\
G\n H4\b L\b                       & 3.13 & -    & 3.13 & -    & 5.70 & -   \\
A\n H4\b H2\b L\b                  & 3.28 & 2.16 & 3.28 & -    & -    & 4.95\\
A\n H4\b L\b                       & 3.50 & -    & 3.50 & -    & -    & 5.28\\
A\n H4\b H2\b G\n L\b              & 2.54 & 1.67 & 2.54 & -    & 4.62 & 3.83\\
A\n H4\b G\n L\b                   & 2.65 & -    & 2.65 & -    & 4.82 & 4.00\\
A\n G\n                            & -    & -    & -    & -    & 5.42 & 4.50\\
H4\n H2\n L\n                      & 4.47 & 2.23 & 4.47 & -    & -    & -   \\
V\b H4\n H2\n L\n                  & 4.24 & 2.12 & 4.24 & 3.04 & -    & -   \\
V\n H4\n H2\n L\n                  & 3.27 & 1.64 & 3.27 & 5.09 & -    & -   \\
\end{tabular}
\end{table}

\begin{table}
\caption{Thresholds to be set for different detectors for coincident
observation of coalescing compact binary sources. A - AIGO, H4 - 4000m
LIGO at Hanford, H2 - 2000m LIGO at Hanford, L - LIGO at Livingston, G
- GEO, V - VIRGO. H - represents both detectors at Hanford
together. For networks with detectors at more than three sites, the
first detector represents the reference detector, whereas the first
three detectors represent the reference 3-detector network, on the
basis of which we calculated the window size ({\it see text}). The
superscripts (\b) and (\n) represent the power and dual recycling
modes of operation respectively.}
\label{t51}
\begin{tabular}{lcccccc}
Network of detectors               & H4   &H2    &L     & V    &G     & A   \\
\hline
H4\b H2\b                          & 6.45 & 3.74 & -    & -    & -    &-    \\
H4\b H2\n                          & 4.37 & 6.03 & -    & -    & -    & -   \\
H4\b L\b or H4\n L\n               & 5.53 & -    & 5.53 &-     & -    & -   \\
H4\b H2\b L\b                      & 5.02 & 2.91 & 5.02 & -    & -    & -   \\
H4\b H2\n L\b                      & 3.87 & 5.34 & 3.87 & -    & -    & -   \\
V\b H4\b H2\b L\b                  & 4.04 & 2.34 & 4.04 & 5.05 & -    & -   \\
V\b H4\b H2\n L\b                  & 3.40 & 4.70 & 3.40 & 4.25 & -    & -   \\
V\b H4\b L\b                       & 4.30 & -    & 4.30 & 5.37 & -    & -   \\
G\n H4\b H2\b L\b                  & 4.43 & 2.57 & 4.43 & -    & 4.21 & -   \\
G\n H4\b H2\n L\b                  & 3.63 & 5.00 & 3.63 & -    & 3.45 & -   \\
G\n H4\b L\b                       & 4.76 & -    & 4.76 & -    & 4.52 & -   \\
A\n H4\b H2\b L\b                  & 4.66 & 2.70 & 4.66 & -    & -    & 3.68\\
A\n H4\b H2\n L\b                  & 3.77 & 5.20 & 3.77 & -    & -    & 2.98\\
A\n H4\b L\b                       & 5.02 & -    & 5.02 & -    & -    & 3.96\\
V\b H4\b H2\b L\b G\n              & 3.62 & 2.10 & 3.62 & 4.52 & 3.44 & -   \\
V\b H4\b H2\n L\b G\n              & 3.14 & 4.33 & 3.14 & 3.91 & 2.98 & -   \\
V\b H4\b L\b G\n                   & 3.80 & -    & 3.80 & 4.75 & 3.62 & -   \\
A\n V\b H4\b H2\b L\b              & 3.80 & 2.20 & 3.80 & 4.75 & -    & 3.00\\
A\n V\b H4\b H2\n L\b              & 3.27 & 4.51 & 3.27 & 4.09 & -    & 2.58\\
A\n V\b H4\b L\b                   & 4.01 & -    & 4.01 & 5.01 & -
& 3.17\\                 A\n H4\b H2\b G\n L\b              & 4.10 &
2.38 & 4.1 0 & -    & 3.89 & 3.24\\
A\n H4\b H2\n G\n L\b              & 3.46 & 4.77 & 3.46 & -    & 3.28 & 2.73\\
A\n H4\b G\n L\b                   & 4.36 & -    & 4.36 & -    & 4.14 & 3.44\\
A\n V\b H4\b H2\b L\b G\n          & 3.42 & 1.98 & 3.42 & 4.27 & 3.25 & 2.70\\
A\n V\b H4\b H2\n L\b G\n          & 3.00 & 4.14 & 3.00 & 3.75 & 2.85 & 2.37\\
A\n V\b H4\b L\b G\n               & 3.57 & -    & 3.57 & 4.46 & 3.39 & 2.82\\
A\n G\n                            & -    & -    & -    & -    & 6.14 & 5.10\\
H4\n H2\n L\n                      & 5.13 & 2.55 & 5.13 & -    & -    & -   \\
V\n H4\n H2\n L\n                  & 4.83 & 2.40 & 4.83 & 3.35 & -    & -   \\
A\n V\b H4\b H2\b L\b G\n          & 4.69 & 2.33 & 4.69 & 3.26 & 1.60
& 1.33 \\
\end{tabular}
\end{table}

\begin{table}
\caption{The range of observation and the source rate of coalescing
binaries for some interesting networks of the initial detectors coming
up at the beginning of the next century. These values correspond to
binaries of chirp mass 1.2$M_\odot$. The calculation of the source
rates is based a conservative estimate of the expectation rate
\{$8\times 10^{-8}$\} Mpc$^{-3}$ yr$^{-1}$ and a maximum limit of
\{$6\times 10^{-5}$\} Mpc$^{-3}$ yr$^{-1}$. For other values of chirp
mass and source rate, see the text. $^\dagger$ Although the last entry
of this table corresponds to a coincident experiment in which all
detectors are operated in dual recycling mode, actually, such an
experiment would not be a meaningful one because the thresholds for
GEO and AIGO (with their present design parameters) are to be
set at very low values compared to LIGOs and VIRGO ({\it for more
explanation see the text}). }
\label{t53}
\begin{tabular}{lccc}
Network of detectors       & Range, ${\cal R}_{90\%}$ &Source rate (yr$^{-1}$)
&Source rate (yr$^{-1}$) \\
                           & (Mpc)  &(conservative estimate) &
(maximum limit)\\
\hline
L\b or H4\b  alone         &37.67     &0.004  &3.19   \\
H4\b H2\b L\b              &57.41     &0.015  &11.30  \\
H4\b H2\n L\b              &74.47     &0.033  &24.67  \\
V\b H4\b H2\b L\b          &68.95     &0.026  &19.58  \\
V\b H4\b H2\b L\b G\n      &77.48     &0.037  &27.78  \\
A\n V\b H4\b H2\b L\b      &73.71     &0.032  &23.92  \\
A\n V\b H4\b H2\b L\b G\n  &82.35     &0.045  &33.35  \\
A\n V\b H4\b H2\n L\b G\n  &94.19     &0.067  &49.91  \\
L\n or H4\n  alone         &104.74    &0.092  &68.63  \\
H4\n L\n                   &144.89    &0.242  &181.68 \\
H4\n H2\n L\n              &156.19    &0.303  &227.58 \\
V\n H4\n H2\n L\n          &165.89    &0.364  &272.68 \\
A\n V\n H4\n H2\n L\n G\n${ }^\dagger$  &170.84    &0.397  &297.83 \\
\end{tabular}
\end{table}


\begin{references}
\bibitem{Web} J. Weber, Phys. Rev., {\bf 117}, 306 (1960)
\bibitem{Kip} K.S. Thorne, in {\it 300 Years of Gravitation}, eds. S. hawking
  	and W. Israel (Cambridge Univ. Press, Cambridge, 1987).
\bibitem{Kip1} K.S. Thorne, in the {\it Proceedings of the Snowmass '95 Summer
	Study on Particle and Nuclear Astrophysics and Cosmology}, eds. E.W.
	Kolb and R. Peccei (World Scientific, Singapore, 1995);
        K.S. Thorne, in {\it Proc. of IAU Symposium 165, Compact Stars in
	Binaries}, eds. J. van Paradijs, E. van den Heuvel, E. Kuulkers (Kluwer
 	Academic Publishers, Dordrecht, The Netherlands, 1995)
\bibitem{Ab} A. Abramovici, W.E. Althouse, R.W.P. Drever, Y. G\"ursel, S.
	Kawamura, F.J. Raab, D. Shoemaker, L. Sievers, R.W. Spero, K.S. Thorne,
 	R.E. Vogt, R. Weiss, S.E. Whitcomb, and M.E. Zucker, Science,
{\bf 256},
        325 (1992).
\bibitem{Virgo} A. Giazotto {\it et al}, {\it The VIRGO project} (INFN, 1989);
        {\sl VIRGO: Final conceptual design} (1992).
\bibitem{Geo} J. Hough {\it et al},{\it Proposal for a Joint German-British
 	Interferometric Gravitational Wave Detector} (1989).
\bibitem{Phin} R. Narayan, T. Piran, and A. Shemi, Astrophy. J. {\bf 379}, L17
	(1991); E.S. Phinney, {\it ibid.} {\bf 380}, L17 (1992).
\bibitem{Ccb} A. Krolak, in {\it Gravitational Wave Data Analysis}, ed. B.F.
	Schutz (Kluwer, Dordrecht, 1989) pp. 59-69; S.V. Dhurandhar,
A. Krolak,
	B. Schutz and J. Watkins, `Filtering the gravitational Wave
Signal of a
	Coalescing Binary', report (unpublished); L.S. Finn, Phys. Rev. D, {\bf
 	46}, 5236(1992)
\bibitem{BSS} B.S. Sathyaprakash and S.V. Dhurandhar, Phys. Rev. D {\bf 44},
 	3819 (1991)
\bibitem{SVD} S.V. Dhurandhar and B.S. Sathyaprakash, Phys. Rev. D {\bf
        49}, 1707 (1994)
\bibitem{FCher} L.S. Finn and D.F. Chernoff, Phys. Rev. D {\bf 47}, 2198 (1993)
\bibitem{KLM} A. Krolak, J.A. Lobo, B.J. Meers, Phys. Rev. D {\bf 48}, 3451
	(1993)
\bibitem{K} {\sl Errors in a few equations given in Ref.\cite{KLM} have been
 	corrected in} A. Krolak, {\it unpublished note} (1994).
\bibitem{Schu2} B.F. Schutz, in {\it Gravitational Collapse and Relativity},
	eds. H. Sato and T. Nakamura (World Scientific, Singapore, 1986),
	pp.350-458.
\bibitem{Schu3} B.F. Schutz, Nature, {\bf 323}, 310 (1986),
\bibitem{Schu4} B.F. Schutz, Cl. Qu. Grav., {\bf 6}, 1761 (1989),
\bibitem{KS} A. Krolak and B.F. Schutz, Gen. Rel. Grav., {\bf 19}, 1163 (1987).
\bibitem{DhTin} S.V. Dhurandhar and M. Tinto, Mon. Not. R. Astron. Soc., {\bf
	 234}, 663 (1988); M. Tinto and S.V. Dhurandhar, Mon. Not. R. Astron.
	 Soc., {\bf 236}, 621 (1989)
\bibitem{Gur} Y. G\"ursel and M. Tinto, Phys. Rev D {\bf 40}, 3884 (1989).
\bibitem{Jar} P. Jaranowski and A. Krolak, Phys. Rev. D {\bf 49}, 1723 (1994).
\bibitem{Vschu} S. Bourzeix, G. Boursin, B. Linet, Phys. Lett. A {\bf 151}, 371
	 (1990).
\bibitem{CutF} C. Cutler and E. Flanagan, Phys. Rev. D {\bf 49}, 2658 (1994)
\bibitem{Mark} D. Markovi\'c, Phys. rev. D {\bf 48}, 4738 (1993)
\bibitem{Shu} B.F. Schutz (1989) {\sl Data analysis requirements of networks of
 	detectors} in {\it Gravitational wave data analysis} Ed. B.F. Schutz
	(Kluwer Academic Publishers, Dordrecht, The Netherlands, 1989) p. 315.
\bibitem{San} J. Sandeman and D.E. McClelland, {\it private communication}.
\bibitem{Wink} W. Winkler, in: The detection of gravitational waves, ed. D.G.
	 Blair (Cambridge University Press, Cambridge, 1991);
\bibitem{Dre} R.W.P. Drever,  in: The detection of gravitational
waves, ed. D.G.
	 Blair (Cambridge University Press, Cambridge, 1991);
\bibitem{Rec} R.W.P. Drever, in: Gravitational Radiation,
ed. N. Deruelle and T.
         Piran (North-Holland, Amsterdam, 1983); P.Fritschel, D. Shoemaker and
         R. Weiss, Ap. Optics, {\bf 31}, 1412 (1992).
\bibitem{FPDL} J.-Y. Vinet, B.J. Meers, C.N. Man, A. Brillet,
Phys. Rev., D {\bf
	 38}, 433 (1988); A. Brillet, J. Gea-Banacloche, G. Leuches, C.N. Man,
	 J.-Y. Vinet, in: The detection of gravitational waves, ed. D.G. Blair
	 (Cambridge University Press, Cambridge, 1991)
\bibitem{Vin} J.-Y. Vinet, B.J. Meers, C.N. Man, A. Brillet, Phys. Rev., D
        {\bf 38}, 433
        (1988); A. Giazotto, Phys. Rep., {\bf 182}, 365 (1989).
\bibitem{Dual} B.J. Meers, Phys. Rev., D {\bf 38}, 2317 (1988); {\it ibid}
        Phys. Lett., A {\bf 142}, 465 (1989); K.A. Strain and B.J. Meers,
        Phys. Rev. Lett., {\bf 66}, 1391 (1991).
\bibitem{Saul90} P.R. Saulson, Phys. Rev., D {\bf 42}, 2437, (1990)
\bibitem{GR} A. Gillespie and F. Raab, Phys. Lett., A {\bf 178}, 357 (1993).
\bibitem{GR95} A. Gillespie and F. Raab, Phys. Rev., D {\bf 52}, 577 (1995).
\bibitem{mod} R. Weiss, Quart. Prog. Rep. Res. Lab. Electronics, MIT {\bf 105},
         54 (1972); C.N. Man, D. Shoemaker, M. Pham Tu, D. Dewey, Phys. Lett.,
         A {\bf 148}, 8 (1990).
\bibitem{Saul84} P.R. Saulson, Phys. Rev., D {\bf 30}, 732 (1984)
\bibitem{VIN} J.-Y. Vinet, {\it private communication},
\bibitem{Win} W. Winkler, {\it private communication},
\bibitem{FU} M. Fujimoto, {\it private communication},
\bibitem{dhs} However, the final choice of this value will crucially depend on
        the mirror losses and the issue is not settled yet
(D.H. Shoemaker, {\it
        private communication}).
\bibitem{bw} We may note here that the conclusion drawn by Krolak, Lobo and
  	 Meers\cite{KLM} that the narrow-band dual recycling with a bandwidth
	 of 6Hz around 100Hz is better than standard recycling for the
	observation of the inspiral waveform from binary systems is actually
        not valid when thermal noise is also taken into account, which they
  	completely neglected. Our numerical calculation shows that actually the
        SNR for initial LIGO reduces by about 39\% as compared to the power
         recycling case.
\bibitem{Tin} M. Tinto and B.F. Schutz, Mon. Not. R. Astron. Soc., {\bf 224},
	 131 (1987); M. Tinto, {\it ibid}, {\bf 226}, 829 (1987); M. Tinto, in
	 {\it Gravitational Wave Data Analysis}, ed. B.F. Schutz (Kluwer,
	 Dordrecht, 1989) p. 299.
\bibitem{ShDh} S. V. Dhurandhar and B.F. Schutz, Phys. Rev. D {\bf 50}, 2390
	 (1994)
\bibitem{BSD} R. Balasubramanium, B.S. Sathyaprakash and S.V. Dhurandhar,
         {\sl Gravitational waves from coalescing binaries : Detection
         strategies and Monte Carlo simulation for parameter estimation} {\it
	 submitted for publication}.
\bibitem{Chris} N. Christensen, Phys. Rev., D {\bf 46}, 5250 (1992).
\bibitem{FLA} E.E. Flanagan, Phys. Rev., D {\bf 48}, 2389 (1993).

\end{references}
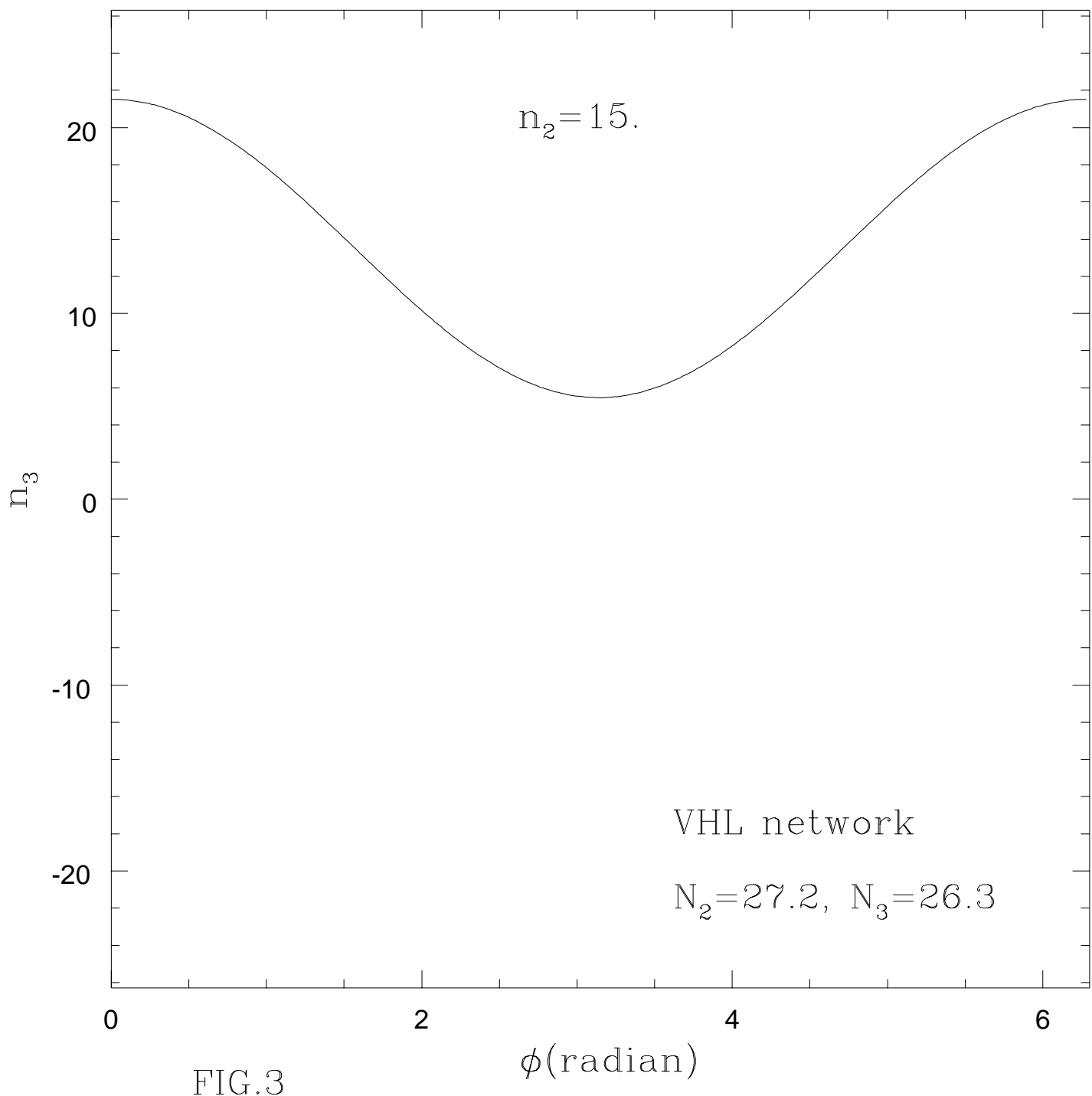
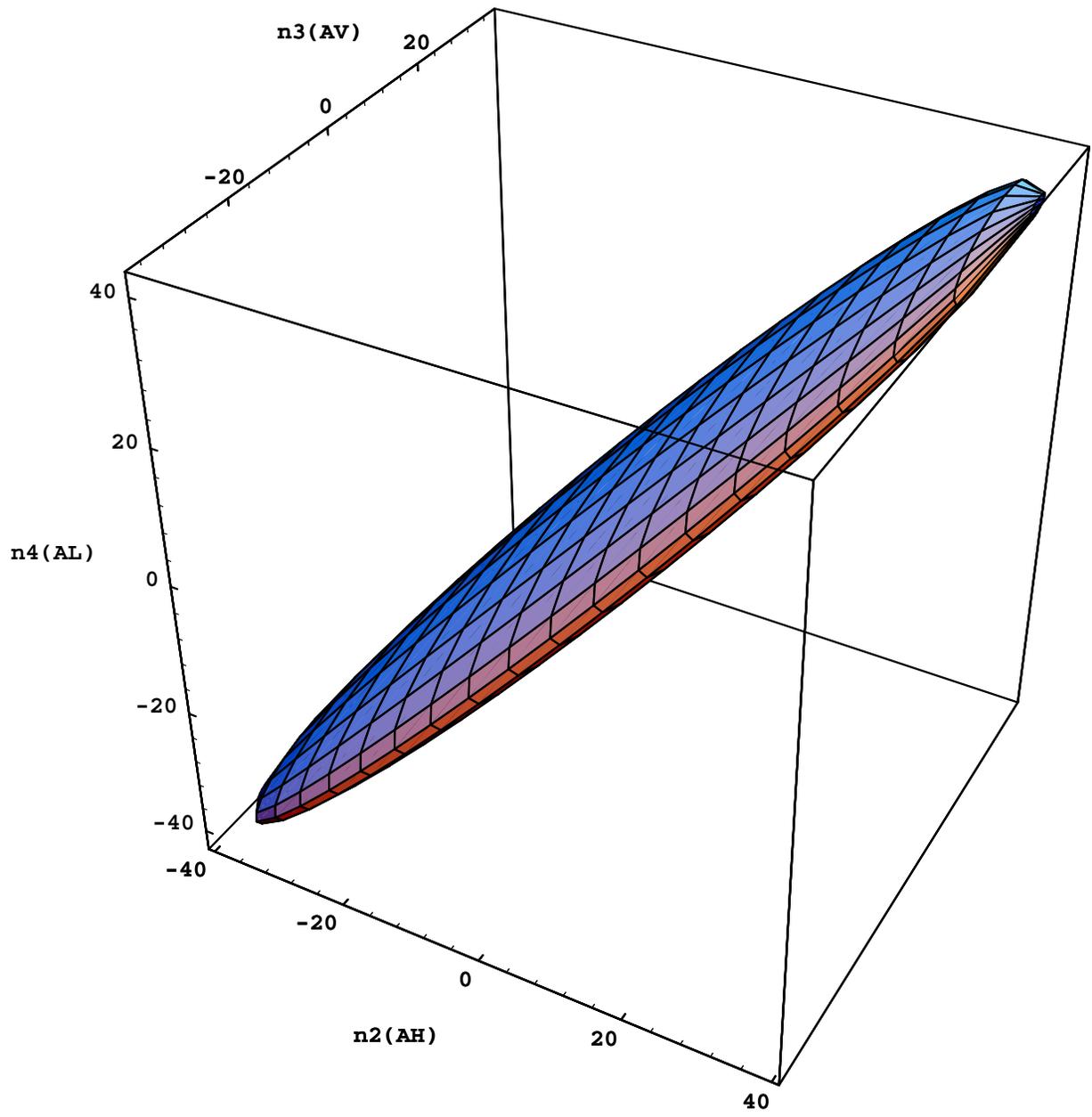
\end{document}